\documentclass[a4paper,11pt]{article}
\usepackage{pos}
\usepackage{diagbox}

\title{Doubly heavy hadron production in ultraperipheral collisions}
\author*[a]{Jun Jiang}
\author[b]{Hao Yang}
\author[a]{Xiao Liang}
\author[a]{Zong-Guo Si}
\author[c,d]{Cong-Feng Qiao}
\author[b,e]{Bing-Wei Long}
\author[a]{Yan-Rui Liu}
\author[a]{Shi-Yuan Li}

\affiliation[a]{School of Physics, Shandong University,\\
Jinan, Shandong 250100, China}

\affiliation[b]{College of Physics, Sichuan University,\\
Chengdu, Sichuan 610065, China}

\affiliation[c]{School of Physical Sciences, University of Chinese Academy of Sciences,\\
Beijing 100049, China}

\affiliation[d]{CAS Key Laboratory of Vacuum Physics,\\
Beijing 100049, China}

\affiliation[e]{Southern Center for Nuclear-Science Theory, Institute of Modern Physics, Chinese Academy of Sciences, \\
Huizhou 516000, Guangdong, China}

\emailAdd{jiangjun87@sdu.edu.cn}
\abstract{The inclusive production of pseudoscalar heavy quarkonia ($\eta_c,\, \eta_b,\, B_c$), double heavy baryons $\Xi_{QQ^\prime}$ ($Q^{(\prime)}=c,\,b$ quarks) and tetraquarks $T_{QQ}$ in heavy ion ultraperipheral collisions (UPCs) is studied.
Numerical results indicate that the experimental investigation of $\eta_c,\, \Xi_{cc}$, and $T_{cc}$ is feasible at the upcoming HL-LHC and future FCC.
Heavy ion UPCs open another avenue for studying the production of these doubly heavy hadrons.}

\FullConference{The 21st International Conference on Hadron Spectroscopy and Structure (HADRON2025)\\
 27 - 31 March, 2025\\
Osaka University, Japan\\}

\begin{document}
\maketitle

\section{Why ultraperipheral collisions?}

In energetic heavy ion collisions, the large proton charge number $Z$ makes highly relativistic ions intense sources of electromagnetic radiation. Within the equivalent photon approximation \cite{vonWeizsacker:1934nji,Williams:1934ad}, this radiation can be treated as fluxes of quasi-real photons.
Thus, heavy ion collisions enable the study of photon-photon, photon-proton, and photon-Pb interactions in ultraperipheral collisions (UPCs) \cite{Baur:2001jj,Bertulani:2005ru,Baltz:2007kq}. In this article, we focus on the photon-photon fusion (photon-initiated) mechanism in heavy ion UPCs.
Compared with central heavy ion collisions, quasi-real photon interactions in UPCs exhibit low event multiplicity. This is because the ion impact parameter exceeds twice the ion radius, keeping the ions intact and eliminating noisy hadron backgrounds.
Compared with photon-photon fusion in proton-proton collisions, the photon density for each ion is significantly enhanced by the squared ion charge $Z^2$, leading to an overall $Z^4$ enhancement in the production rate.
Compared with photon-photon interactions at electron-positron colliders, de-excitation photons with energies up to 80 GeV and 600 GeV \cite{Shao:2022cly} can be emitted in Pb-Pb UPCs at the High-Luminosity Large Hadron Collider (HL-LHC) \cite{Bruce:2018yzs,Klein:2020nvu,dEnterria:2022sut} and the Future Circular Collider (FCC) \cite{FCC:2018vvp,Dainese:2016gch}, respectively.
Therefore, photon-initiated UPC processes provide another promising platform for studying the production of heavy hadrons in an electromagnetic environment.

\section{Why doubly heavy hadrons?}

Doubly heavy hadrons—including quarkonia $\mathcal{H}(Q\bar{Q})$, baryons $\Xi_{QQ^{\prime}}$, and tetraquarks $T_{QQ^{\prime}}$ (with $Q^{(\prime)}=c,\,b$ quarks)—are ideal systems for investigating both perturbative and nonperturbative quantum chromodynamics (QCD).
Heavy quarkonia are bound states composed of a heavy quark and its corresponding heavy antiquark. Their production involves two key steps: the perturbative creation of a heavy quark-antiquark pair ($Q\bar{Q}$), followed by the nonperturbative hadronization of this pair into a bound heavy quarkonium state ($\mathcal{H}(Q\bar{Q})$).
The nonrelativistic QCD (NRQCD) factorization formalism \cite{Bodwin:1994jh,Petrelli:1997ge} was developed to study quarkonium production and decay, as it explicitly separates perturbative and nonperturbative effects. The formation of a bound $Q\bar{Q}$ intermediate state with definite $J^{PC}$ and color configuration can be calculated perturbatively, using a double expansion in the strong coupling constant $\alpha_s$ and the relative velocity $v$ between the heavy quark $Q$ and antiquark $\bar{Q}$. By contrast, the nonperturbative hadronization of these intermediate states into physical heavy quarkonia ($\mathcal{H}(Q\bar{Q})$) is described by long-distance matrix elements (LDMEs).
For doubly heavy baryons ($\Xi_{QQ^{\prime}}$) and tetraquarks ($T_{QQ^{\prime}}$), the doubly heavy components within these hadrons typically exhibit nonrelativistic behavior. This causes the heavy quarks to remain tightly bound, forming a heavy-heavy diquark ($QQ^\prime$). As a result, the production of doubly heavy baryons and tetraquarks can also be described within the NRQCD framework, which factorizes the process into two stages:
the perturbative production of a doubly heavy diquark with a specific spin-color structure, and the nonperturbative hadronization of this diquark via QCD mechanisms which is encoded in LDMEs and fragmentation functions.

\section{Heavy quarkonium production in UPCs}

In this section, the inclusive production of pseudoscalar heavy quarkoniua ($\eta_c,~\eta_b$ and $B_c$) via photon-photon fusion in UPCs are calculated to QCD next-to-leading order within NRQCD framework \cite{Jiang:2024vuq}.

Within the equivalent photon approximation (EPA) formalism \cite{vonWeizsacker:1934nji,Williams:1934ad}, the cross section for producing final state $X$ via photon-photon fusion in UPCs can be factorized into a convolution of the equivalent photon spectra and the subprocess cross section for $X$ production:
\begin{equation}
\sigma(\mathrm{A} \mathrm{B} \stackrel{\gamma \gamma}{\longrightarrow} \mathrm{A} \mathrm{B} X)=\int \frac{d x_1}{x_1} \frac{d x_2}{x_2} f(x_1) f(x_2) \times \mathrm{d} \hat{\sigma}({\gamma \gamma \rightarrow X}).
\label{eq:totalXS}
\end{equation}
Here, $x_{i} = E_i/E_{\mathrm{beam}}$ denotes the ratio of the photon energy $E_i$ to the beam energy $E_{\mathrm{beam}}$. The photon spectrum $f(x)$ generated by an ion with charge $Z$ is given by \cite{Jackson:1999,Shao:2022cly}
\begin{equation}
f(x)=\frac{2 \alpha Z^2}{\pi}\left[\chi K_0(\chi) K_1(\chi)-\left(1-\gamma_{\mathrm{L}}^{-2}\right) \frac{\chi^2}{2}\left(K_1^2(\chi)-K_0^2(\chi)\right)\right],
\label{eq:photonpdf}
\end{equation}
where the variable $x$ is incorporated into $\chi \equiv x m_N b_{\mathrm{min}}$. In this expression, $m_N$ is the nucleon mass, and the minimum impact parameter $b_{\mathrm{min}}$ is set to the nuclear radius. $K_0$ and $K_1$ are the modified Bessel functions of the second kind, of order zero and one, respectively. $\gamma_L=E_{\mathrm{beam}}/m_N$ is the Lorentz factor.
For the subprocess of inclusive pseudoscalar heavy quarkonium production via photon-photon fusion—i.e., $\gamma+\gamma\to \mathcal{H}(Q_1\bar{Q}_2)+Q_2+\bar{Q}_1$, where $\mathcal{H}(Q_1\bar{Q}_2) = \eta_c$, $\eta_b$, or $B_c$, and $Q_{1,2}$ correspond to heavy charm or bottom quarks—we have calculated the cross section up to next-to-leading order in QCD. Detailed discussions of this calculation can be found in Ref. \cite{Yang:2022yxb,Jiang:2024vuq}.

%%%%%%%%%%%%%%%%%%%%%%%%%%%%%%%%%%%%%%%%%%%%%%%%%%%%%%%%%
\begin{table}[ht]
    \caption{The LO and NLO total cross sections (in nb) for $\eta_c+c+\bar{c}$, $\eta_b +b+\bar{b}$ and $B_c +b+\bar{c}$ production via photon-photon fusion in ultraperipheral Pb-Pb collision at $\sqrt{S_{\mathrm{NN}}}=$ 5.52 and 39.4 TeV. Here,  scale $\mu=\sqrt{m_H^2 + p_{t}^2}$ with $m_H$ being the mass of heavy quarkonia, and the transverse momentum cut $ 1\ {\rm GeV} \le p_{t} \le 50$ GeV is employed.}
    \begin{center}
       \begin{tabular}{c|c|c|c}
        \hline
        processes  & $\eta_c +c\bar{c}$   &  $\eta_b +b\bar{b}$  & $B_c + b\bar{c}$ \\
        \hline
        $\sigma_{\rm LO}$ ($5.52 $ TeV)&
        $1.7\times 10^2$ &
        $0.034$ &
        $0.57$ \\
        $\sigma_{\rm NLO}$ ($5.52 $ TeV)&
        $1.9\times 10^2$ &
        $0.027$ &
        0.47 \\
        K-factor &
        $1.2$ &
        $0.78$ &
        0.83 \\
        \hline
         $\sigma_{\rm LO}$ ($39.4 $ TeV)&
        $1.1\times 10^3$ &
        $0.46$ &
        6.5 \\
         $\sigma_{\rm NLO}$ ($39.4 $ TeV)&
        $1.3\times 10^3$ &
        $0.30$ &
        5.8\\
         K-factor &
        1.2 &
        $0.66$ &
       0.90 \\
        \hline
      \end{tabular}
    \end{center}
    \label{all process}
\end{table}
%%%%%%%%%%%%%%%%%%%%%%%%%%%%%%%%%%%%%%%%%%%%%%%%%%%%%%%%%

In table \ref{all process}, we present the total cross sections for the inclusive production of pseudoscalar quarkonia ($\eta_c,\, \eta_b,\, B_c$) via photon-photon fusion in Pb-Pb UPCs.
We consider two nucleon-nucleon c.m. energies $\sqrt{S_{\mathrm{NN}}}=$ 5.52 and 39.4 TeV, which are typical collision energies at current LHC and future HL-LHC and FCC.
The K-factor is defined as the ratio of cross sections of NLO to that of LO, $\sigma_{\mathrm{NLO}}/\sigma_{\mathrm{LO}}$.
Numerical results indicate that the NLO corrections are significant.
The NLO corrections improve the cross sections of $\eta_c$ by about 16\% and 21\% for $\sqrt{S_{\mathrm{NN}}}=$ 5.52 and 39.4 TeV, respectively.
While the NLO corrections for both $\eta_b$ and $B_c$ are negative;
cross sections at NLO for $\eta_b$ decrease by 22\% and 43\% for $\sqrt{S_{\mathrm{NN}}}=$ 5.52 and 39.4 TeV respectively, and the two percentages are 17\% and 10\% for $B_c$.
Note that the vector $B_c^*$ signal can't be separated from pseudoscalar $B_c$ meson in experiments and will decay into the later electromagnetically with almost 100\% probability. We obtain the LO cross sections for $B_c^*$ meson as 7.0 $\mathrm{nb}$ and 65 $\mathrm{nb}$ for $\sqrt{S_{\mathrm{NN}}}=$ 5.52 TeV and 39.4 TeV respectively, which are about one order of magnitude greater than those of pseudoscalar $B_c$ accordingly.

We move to discuss the events for such inclusive production in Pb-Pb UPCs.
Taking the integrated luminosity per typical run ${\cal L}_{\mathrm{int}}=5 \, {\mathrm{nb}}^{-1}$ and nucleon-nucleon c.m. energy $\sqrt{S_{\mathrm{NN}}}=5.52$ TeV at HL-LHC \cite{Shao:2022cly}, we roughly have $970^{+500}_{-320}$ events produced for $\eta_c$, where the uncertainties are caused by the charm quark mass $m_c = 1.5 \pm 0.1$ GeV. And there are roughly 0 event for $\eta_b$ and 2 events for $B_c$ meson.
Taking the integrated luminosity per typical run ${\cal L}_{\mathrm{int}}=110 \, {\mathrm{nb}}^{-1}$ and nucleon-nucleon c.m. energy $\sqrt{S_{\mathrm{NN}}}=39.4$ TeV at FCC \cite{Shao:2022cly}, the events produced are roughly $(1.4^{+0.65}_{-0.43}) \times 10^5$, 30 and 640 events for $\eta_c, \, \eta_b$ and $B_c$ mesons, respectively.
So one might observe the $\eta_c$ signals produced via photon-photon fusion in Pb-Pb UPCs at future HL-LHC and FCC experiments.

%%%%%%%%%%%%%%%%%%%%%%%%%%%%%%%%%%%%%%%%%%%%%%%%%%%%%%%%%
\begin{figure}[b]
    \centering
    \includegraphics[width=0.8\textwidth]{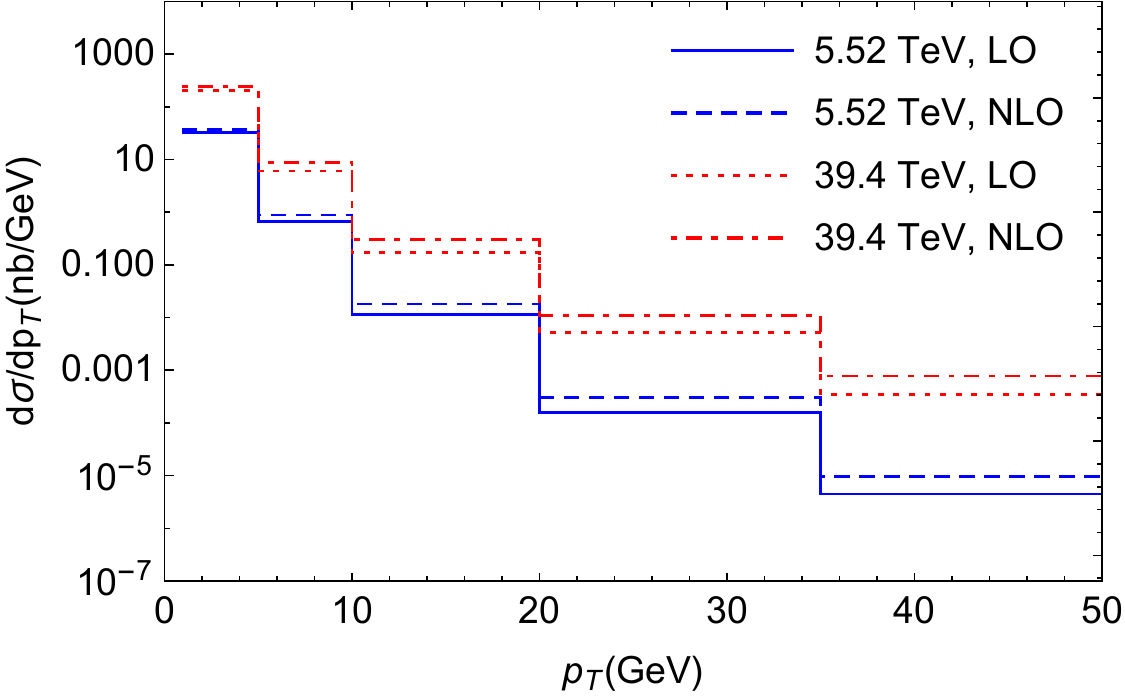}
    \caption{The differential transverse momentum distribution $d\sigma /dp_t$ of $\eta_c$ for the $\eta_c+c+\bar{c}$ production via photon-photon fusion in ultraperipheral Pb-Pb collision at $\sqrt{S_{\mathrm{NN}}}=$ 5.52 and 39.4 TeV. Here, $\mu=\sqrt{4m_c^2 + p_{t}^2}$ and the transverse momentum cut $ 1\ {\rm GeV} \le p_{t} \le 50$ GeV is employed.}
    \label{figptv}
\end{figure}
%%%%%%%%%%%%%%%%%%%%%%%%%%%%%%%%%%%%%%%%%%%%%%%%%%%%%%%%%

We further discuss the uncertainties arising from the renormalization scale $\mu$ at both LO and NLO for $\eta_c+c+\bar{c}$ production. The scale $\mu=r\sqrt{4m_c^2 + p_{t}^2}$ varies by a factor of $r=\{0.5,1,2\}$.
It is found that the cross sections or events at both LO and NLO decrease as the renormalization scale $\mu$ increases.
More specially, the LO estimate at $\sqrt{S_{\mathrm{NN}}}=$ 5.52 TeV increases
by 91\% and decreases by 39\% for $\mu$ varies by a factor of 1/2 and 2, respectively.
At NLO, these two percentages becomes 65\% and 32\%.
At $\sqrt{S_{\mathrm{NN}}}=$ 39.4 TeV, the LO estimate increases
by 90\% and decreases by 38\% for $\mu$ varies by a factor of 1/2 and 2, respectively.
While at NLO, these two percentages becomes 67\% and 32\%.
The NLO correction improve the $\mu$ dependence as we expect.
We also find that the K-factors increase as scale $\mu$ increases, which indicates that the NLO corrections are more remarkable at high scale region.

In figure \ref{figptv}, we present the differential distribution versus the transverse momentum $p_t$ of $\eta_c$ for $\eta_c+c+\bar{c}$ production via photon-photon fusion in Pb-Pb UPCs.
The differential cross sections for both LO and NLO and at both $\sqrt{S_{\mathrm{NN}}}=$ 5.52 TeV and 39.4 TeV decrease monotonically as the transverse momentum $p_t$ of $\eta_c$ grows from 1 to 50 GeV.
Additionally, the positive NLO corrections lead to an increase in amounts in comparison with the LO results in the distributions, while the overall lineshapes are preserved.

\section{Doubly heavy baryon and tetraquark production in UPCs}

We discuss the doubly heavy baryon $\Xi_{QQ'}$ and tetraquark $T_{QQ}$ production through photon-photon fusion and photon-gluon interaction (photoproduction) in UPCs at the LHC and FCC in this section \cite{Yang:2024ysg}.
We adopt a phenomenological diquark fragmentation model to describe the production of baryon $\Xi_{QQ'}$ and compact tetraquark $T_{QQ}$.
This diquark fragmentation contribution becomes dominant only in the large transverse momentum $p_t$ region.
In contrast, there are contributions from other production mechanisms, {\it i.e.} all the quark constitutes are produced at short-distance first then followed by the hadronization. Such contributions would dominate in the relatively small $p_t$ region, which are neglected here.

In the phenomenological diquark fragmentation model, the production of baryon $\Xi_{QQ'}$ and compact tetraquark $T_{QQ}$ has been factored out into two steps: the perturbative production of the $(QQ^{\prime})$-diquark because of the nonrelativistic nature of doubly heavy diquark, and the nonperturbative hadronization of colored $(QQ^{\prime})$-diquark into baryon $\Xi_{QQ'}$ or compact tetraquark $T_{QQ}$.
The creation of $(QQ^{\prime})$-diquark can be related to the production of quark-antiquark pair $(Q\bar{Q}^{\prime})$ by reversing one of the Dirac fermion chain and introducing the color structures of diquarks \cite{Jiang:2024lsr,Yang:2024ysg}.
The hadronization probabilities of $(QQ^{\prime})$-diquark into baryon $\Xi_{QQ'}$ or compact tetraquark $T_{QQ}$ can be determined by fitting the experimental results, the potential models, and the heavy diquark-antiquark symmetry \cite{Jiang:2024lsr,Yang:2024ysg}.
This phenomenological diquark fragmentation model has been widely employed to study the doubly heavy baryons \cite{Ma:2003zk,Jin:2014nva,Li:2020ggh,Sun:2020mvl,Tian:2023uxe,Zhan:2023jfm,Yang:2024ysg} and $T_{cc}$ \cite{Yang:2024ysg,Niu:2024ghc}.

%%%%%%%%%%%%%%%%%%%%%%%%%%%%%%%%%%%%%%%%%%%%%%%%%%%%%%%%%
\begin{table}[htbp!]	
	\caption{The cross sections for $\gamma+\gamma\to \Xi_{cc}[n]+\bar{c}\bar{c}$ through UPCs at the HL-LHC and FCC.}
	\begin{center}
		\begin{tabular}{c|c|c|c|c|c}
			\hline
			Collisions & $\sqrt{s_{NN}}$ (TeV) & $ \Xi_{cc}[{^3S_1}\mbox{-}\bar{\mathbf{3}}] $ & $\Xi_{cc}[{^1S_0}\mbox{-}\mathbf{6}]$ & Total & $N_{\Xi_{cc}}$ \\
			\hline
			Pb-Pb      & 5.52  & 270 nb    & 9.53 nb      & 279.5 nb  &$1.40\times10^3$\\
			\hline
			Xe-Xe      & 5.86  & 65.9 nb   & 2.38 nb      & 68.28 nb  &$2.05\times10^3$\\
			\hline
			Kr-Kr      & 6.46  & 17.8 nb    & 0.663 nb    & 18.46 nb  &$2.21\times10^3$\\
			\hline
			Ar-Ar      & 6.3   & 1.36 nb    & 0.0518 nb   & 1.411 nb  &$1.55\times10^3$\\
			\hline
			Ca-Ca      & 7.0   & 2.31 nb    & 0.0886 nb   & 2.398 nb  &$1.92\times10^3$\\
			\hline
			O-O        & 7.0   & 77.1 pb     & 3.03 pb    & 80.13 pb  &$9.61\times10^2$\\
			\hline
			p-Pb       & 8.8   & 203 pb     & 8.33 pb     & 211.33 pb &$2.11\times10^2$\\
			\hline
			p-p        & 14    & 89.6 fb     & 3.99 fb    & 93.59 fb  &$1.43\times10^4$\\
			\hline
			\hline
			Pb-Pb      & 39.4  & 1780 nb    & 74.5 nb     & 1854 nb   &$2.04\times10^5$\\
			\hline
			p-Pb       & 62.8  & 728 pb    & 32.8 pb      & 760.8 pb  &$2.20\times10^4$\\
			\hline
			p-p        & 100   & 233 fb     & 10.9 fb     & 243.9 fb  &$2.44\times10^5$\\
			\hline		
		\end{tabular}
	\end{center}
	\label{Tabrr2Xicc}
\end{table}
%%%%%%%%%%%%%%%%%%%%%%%%%%%%%%%%%%%%%%%%%%%%%%%%%%%%%%%%%

We first discuss the production rates for $\gamma+\gamma\to \Xi_{cc}[n]+\bar{c}\bar{c}$ in UPCs at the HL-LHC and FCC, where $n$ stands for the spin-color configurations ${^3S_1}\mbox{-}\bar{\mathbf{3}}$, ${^1S_0}\mbox{-}\mathbf{6}$ {\it etc}.
The cross sections for each spin-color state of $\Xi_{cc}$ are listed in table \ref{Tabrr2Xicc}.
The contribution from $[{^1S_0}\mbox{-}\mathbf{6}]$ configuration is only $3\% \sim 5\%$ to that of $[{^3S_1}\mbox{-}\bar{\mathbf{3}}]$, the ratio holds for $\Xi_{bb}$.
With the designed integrated luminosities and aggregating the contributions of diquark in all spin-color configurations, the produced $\Xi_{cc}$ numbers via various UPCs at the HL-LHC are around $10^{3}$.
As the collision energies and luminosities are highly improved at the FCC, the yields for $\Xi_{cc}$ can be increased by one or two orders of magnitude.
Further considering the relative possibilities for various light quarks of $u:d:s \sim 1:1:0.3$, and the brancing fractions of reconstruction channels $Br(\Xi_{cc}^{++}\to\Lambda_c^+ K^-\pi^+\pi^+) \approx 10\%$ \cite{Yu:2017zst}, $Br(\Lambda_c^+\to pK^+\pi^+)\approx 5\%$ \cite{LHCb:2013hvt}, the observations of $\Xi_{cc}^{++}$ via photon-photon fusion in UPCs may be expected at the future FCC.
We also study the production rates for $\Xi_{bc/bb}$ via photon-photon fusion in UPCs at HL-LHC and FCC, and the cross sections are too small to carry out any experimental observations \cite{Yang:2024ysg} and we do not discuss them here.

%%%%%%%%%%%%%%%%%%%%%%%%%%%%%%%%%%%%%%%%%%%%%%%%%%%%%%%%%
\begin{table}[ht]
	\caption{The cross sections for $\Xi_{cc}[{^3S_1}\mbox{-}\bar{\mathbf{3}}]$ ($\Xi_{cc}[{^1S_0}\mbox{-}\mathbf{6}]$) (in unit of nb) under different $m_c$ and renormalization scales in ultraperipheral Pb-Pb collision at 5.52 TeV.}
	\begin{center}
			\begin{tabular}{|l|c|c|c|}
				\hline
				\diagbox{$\mu$ (GeV)}{$m_c$ (GeV)} & \hspace{1.cm}1.7\hspace{1.cm} & \hspace{1.cm}1.8\hspace{1.cm} & \hspace{1.cm}1.9\hspace{1.cm} \\
				\hline
				$\dfrac{1}{2}\sqrt{4m_c^2+p_T^2}$ & 752 (25.8) & 496 (16.9) & 334 (11.4) \\
				\hline
				$\sqrt{4m_c^2+p_T^2}$             & 404 (14.3) & 271 (9.53) & 185 (6.48) \\
				\hline
				$2\sqrt{4m_c^2+p_T^2}$            & 252 (9.09) & 170 (6.10) & 117 (4.18) \\
				\hline		
			\end{tabular}
	\end{center}
	\label{TabUncertainty}
\end{table}
%%%%%%%%%%%%%%%%%%%%%%%%%%%%%%%%%%%%%%%%%%%%%%%%%%%%%%%%%

%%%%%%%%%%%%%%%%%%%%%%%%%%%%%%%%%%%%%%%%%%%%%%%%%%%%%%%%%
\begin{figure}[htbp!]			
	\centering
	\includegraphics[width=0.49\textwidth]{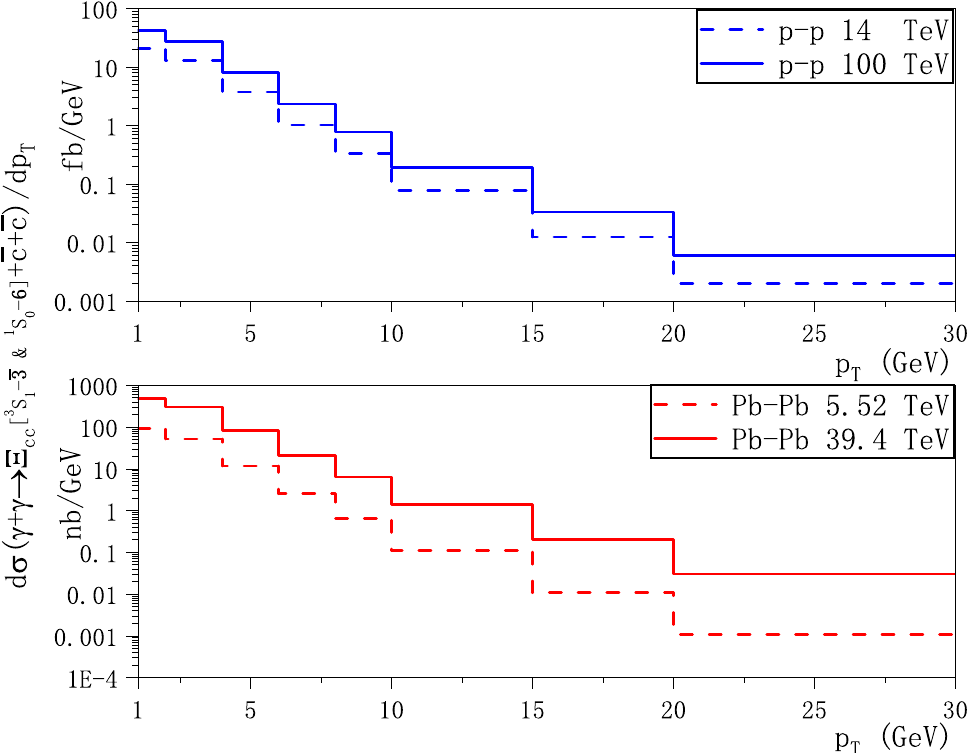}
	\includegraphics[width=0.49\textwidth]{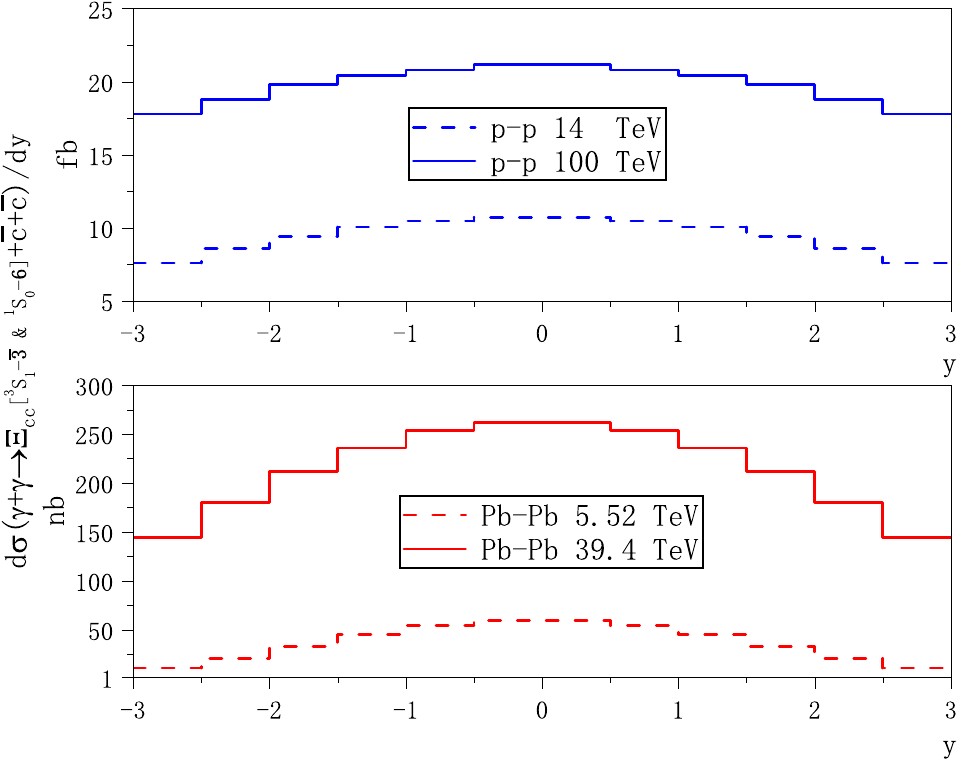}
	\caption{The transverse momentum $p_T$ and rapidity y distributions for $\Xi_{cc}$ production via photon-photon fusion in UPCs. Here, for the $p_T$ distribution, y is cut to be $[-3,3]$; for the y distribution, $p_T$ is cut to be $1\mbox{-}30$ GeV.}
	\label{Figrr2Xiccpty}
\end{figure}
%%%%%%%%%%%%%%%%%%%%%%%%%%%%%%%%%%%%%%%%%%%%%%%%%%%%%%%%%

We further perform an elaborate phenomenological analysis on the subprocess $\gamma+\gamma\to \Xi_{cc}[{^3S_1}\mbox{-}\bar{\mathbf{3}} + {^1S_0}\mbox{-}\mathbf{6}] + \bar{c}\bar{c}$ in UPCs.
To estimate the theoretical uncertainties caused by the charm quark mass and renormalization scale, we set the charm mass to be 1.7, 1.8 and 1.9 GeV, and the renormalization scales are chosen to be $\dfrac{1}{2}$, 1 and 2 times the $\Xi_{cc}$ transverse mass, and the results as shown in table \ref{TabUncertainty}.
The transverse momentum and rapidity distributions of $\Xi_{cc}$ in UPCs are given in figure \ref{Figrr2Xiccpty}, with the $p_T = 1-30$ GeV and $y=[-3,3]$.
The cross sections decrease rapidly versus high $p_T$, showing a logarithmic dependence of $p_T$.
As the UPC events are characterized by a large rapidity gap between the produced particles and the interacting nucleus accompanied by forward neutron emission from the de-excitation of nucleus \cite{Baltz:2007kq}, the resulting rapidity distribution is relative narrow and centered at midrapidity.

%%%%%%%%%%%%%%%%%%%%%%%%%%%%%%%%%%%%%%%%%%%%%%%%%%%%%%%%%
\begin{table}[htbp!]
	\caption{The cross sections for photoproduction $g+\gamma\to \Xi_{cc}[n]+\bar{c}\bar{c}$ at the HL-LHC and FCC. The cross sections in brackets are the contributions from $\gamma+g$ channel, which are different from $g+\gamma$ channel in the p-Pb collision; while the two contributions are absolutely equal for same ion collision. The total cross sections contains both two contributions.}
	\begin{center}
			\begin{tabular}{c|c|c|c|c|c}
				\hline
				Collisions & $\sqrt{s_{NN}}$\ (TeV) & $ \Xi_{cc}[{^3S_1}\mbox{-}\bar{\mathbf{3}}] $ & $\Xi_{cc}[{^1S_0}\mbox{-}\mathbf{6}]$ & Total & $N_{\Xi_{cc}}$ \\
				\hline
				Pb-Pb      & 5.52  & 82.9 $\mu$b         & 7.09 $\mu$b          & 179.98 $\mu$b &$9.00\times10^5$\\
				\hline
				Xe-Xe      & 5.86  & 24.9 $\mu$b         & 2.14 $\mu$b          & 54.08 $\mu$b  &$1.62\times10^6$\\
				\hline
				Kr-Kr      & 6.46  & 8.34 $\mu$b         & 0.72 $\mu$b          & 18.12 $\mu$b  &$2.17\times10^6$\\
				\hline
				Ar-Ar      & 6.3   & 1.08 $\mu$b         & 0.094 $\mu$b         & 2.35 $\mu$b   &$2.58\times10^6$\\
				\hline
				Ca-Ca      & 7.0   & 1.46 $\mu$b         & 0.127 $\mu$b         & 3.17 $\mu$b   &$2.53\times10^6$\\
				\hline
				O-O        & 7.0   & 108 nb              & 9.44 nb              & 234.88 nb     &$2.81\times10^6$\\
				\hline
				p-Pb       & 8.8   & 628 (44.2) nb       & 54.2 (4.01) nb       & 730.41 nb     &$7.30\times10^5$\\
				\hline
				p-p        & 14    & 325 pb              & 29.6 pb              & 709.2 pb      &$1.06\times10^8$\\
				\hline
				\hline
				Pb-Pb      & 39.4  & 374 $\mu$b          & 34.3 $\mu$b          & 816.6 $\mu$b  &$8.98\times10^7$\\
				\hline
				p-Pb       & 62.8  & 2.75 (0.14) $\mu$b  & 0.25 (0.013) $\mu$b  & 3.15  $\mu$b  &$9.13\times10^7$\\
				\hline
				p-p        & 100   & 1090 pb             & 103 pb               & 2386 pb       &$2.38\times10^9$\\
				\hline		
			\end{tabular}
	\end{center}
	\label{Tabgr2Xicc}
\end{table}
%%%%%%%%%%%%%%%%%%%%%%%%%%%%%%%%%%%%%%%%%%%%%%%%%%%%%%%%%

\begin{figure}[htbp!]			
	\centering
	\includegraphics[width=0.49\textwidth]{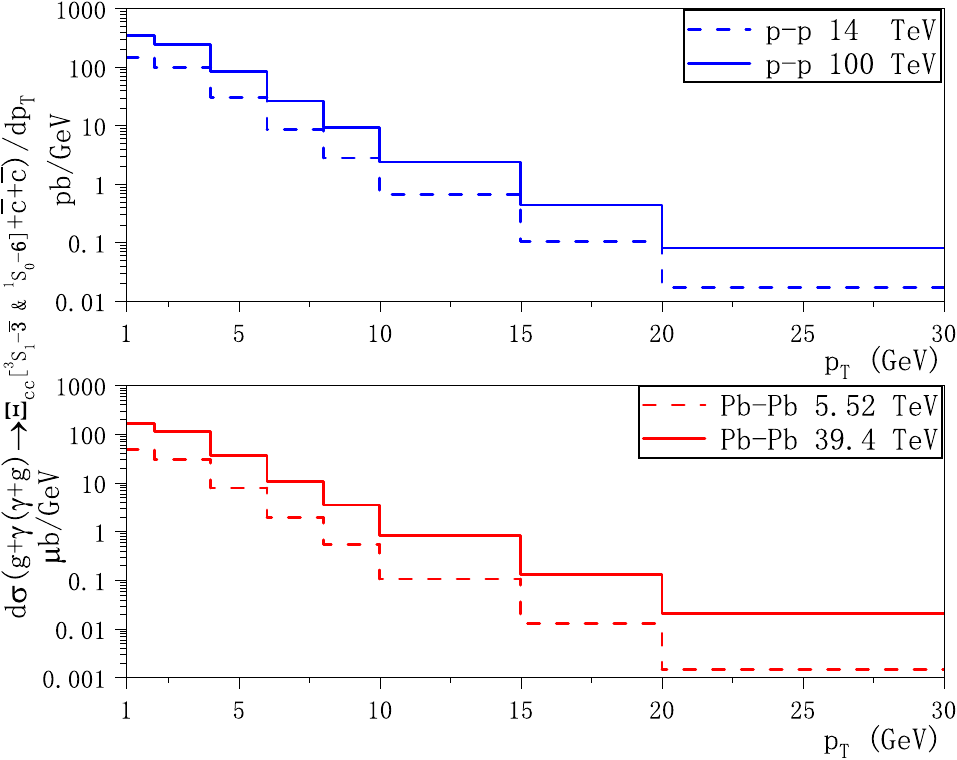}
	\includegraphics[width=0.49\textwidth]{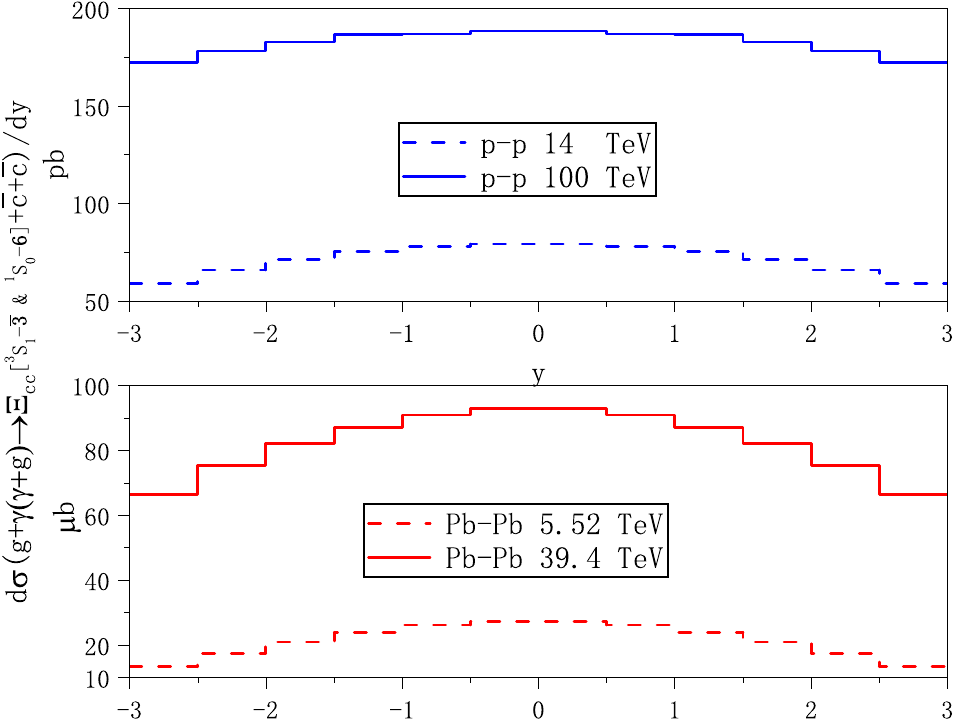}
	\caption{The transverse momentum $p_T$ and rapidity y distributions for $\Xi_{cc}$ production via photoproduction in UPCs. Here, for the $p_T$ distribution, y is cut to be $[-3,3]$; for the y distribution, $p_T$ is cut to be $1\mbox{-}30$ GeV.}
	\label{Figgr2Xiccpty}
\end{figure}

We then discuss the photoproduction subprocess, $g+\gamma \to \Xi_{cc}[n]+\bar{c}\bar{c}$ in UPCs.
The cross sections for two spin-color configurations and the events under various ion collision systems at HL-LHC and FCC are presented in table \ref{Tabgr2Xicc}.
In comparison with photon-photon fusion, the ratio by two spin-color configurations $\sigma(\Xi_{cc}[{^1S_0}\mbox{-}\mathbf{6}])/\sigma(\Xi_{cc}[{^3S_1}\mbox{-}\bar{\mathbf{3}}])$ increases to $8\% \sim 9\%$.
Considering the reconstruction channels $\Xi_{cc}^{++}\to\Lambda_c^+ K^-\pi^+\pi^+$, $\Lambda_c^+\to pK^+\pi^+$, the events for $\Xi_{cc}^{++}$ through such reconstruction channels are $1.95\times10^{3}$ for Pb-Pb collision, and $2.3\times10^{5}$ for p-p collision at HL-LHC.
As the collision energies and luminosities are highly improved at the FCC, the yields for $\Xi_{cc}$ can be increased by one or two magnitudes.
As the physical potential to observe doubly charmed baryon via photoproduction channel is large, we further perform the transverse momentum and rapidity distributions of $\Xi_{cc}$ in figure \ref{Figgr2Xiccpty}, with the $p_T$ is cut to be $1\mbox{-}30$ GeV and y is set to be $[-3,3]$.

%%%%%%%%%%%%%%%%%%%%%%%%%%%%%%%%%%%%%%%%%%%%%%%%%%%%%%%%%
%%%%%%%%%%%%%%%%%%%%%%%%%%%%%%%%%%%%%%%%%%%%%%%%%%%%%%%%%

We also study the photoproduction for $\Xi_{bc/bb}$ via various ion collision systems at HL-LHC and FCC.
As the heavy constituent quarks for $bc$-diquark are different, the asymmetry by exchanging identical heavy quark is not hold, hence all diquark configurations $bc[{^3S_1}\mbox{-}\bar{\mathbf{3}}],\ bc[{^3S_1}\mbox{-}\mathbf{6}],\ bc[^1S_0\mbox{-}\bar{\mathbf{3}}]$, and $bc[^1S_0\mbox{-}\mathbf{6}]$ will contribute and their contributions are comparable.
At the HL-LHC, $10^{4}\sim 10^{6}$ $\Xi_{bc}$ could be produced, which increase to $2\times10^{6}\sim 7\times10^{7}$ at FCC, lead to open possibility for experimental researches. For the $\Xi_{bb}$, the cross sections are further suppressed, hence the experimental investigation for $\Xi_{bb}$ through ion-ion photoproduction may be not feasible.

Considering the number of produced $(cc)$-diquark is large, the compact doubly heavy tetraquark $T_{cc}$ which is composed of $cc$-diquark and light antidiquark $\bar{u}\bar{d}$ can also be produced.
In the diquark-antiquark symmetry, $cc$-diquark cluster in color anti-triplet $\bar{\mathbf{3}}$ may served as heavy antiquark.
Therefore, in the heavy quark limit, the fragmentation probability for $(cc)_{\bar{\mathbf{3}}}\to T_{cc}(cc\bar{u}\bar{d})$ can be approximately described by the probability of $\bar{c}_{\bar{\mathbf{3}}} \to \bar{\Lambda}_c^-(\bar{c}\bar{u}\bar{d})$.
The transition probability is described by the fragmentation function $D_{\Lambda_c/c}(z)$ of charm quark to charm baryon.
The fragmentation fraction for $c\to \Lambda_c^+$ is measured to be $20.4\%$ at the LHC \cite{ALICE:2021dhb}.
In this way, the production cross sections for $T_{cc}^+$ can be estimated by $\sigma(\Xi_{cc})\times 25.6\%$.
$T_{cc}^{+}$ can be reconstructed by $D^0D^0\pi^+$ channel (supposed to be $100\%$), with subsequent $D^0$ decay branching fraction $Br(D^0 \to K^-\pi^+) = 3.94\%$ \cite{ParticleDataGroup:2022pth}.
Then for photon-photon fusion, the expected yields are 81 events for Pb-Pb collision and 96 for pp collision at future FCC, but unexpected at HL-LHC.
For photoproduction, they are 357 events for Pb-Pb and $4\times 10^4$ for pp at HL-LHC, and higher at FCC.

\section{Summary}

We investigate the inclusive production of pseudoscalar heavy quarkonia via photon-photon fusion in ultraperipheral collisions (UPCs) within the framework of NRQCD.
Additionally, we study the inclusive production of doubly heavy baryons and tetraquarks in UPCs—via both photon-photon fusion and photoproduction—using a phenomenological diquark fragmentation model.
We find that $\eta_c$ and $\Xi_{cc}^{++}$ could potentially be observed via photon-photon fusion in UPCs at the HL-LHC and future FCC, respectively. For photoproduction in UPCs, the production yield of $\Xi_{cc}^{++}$ is higher than that from photon-photon fusion. Furthermore, studies of $T_{cc}$ via photoproduction in UPCs could be carried out at both the HL-LHC and future FCC.
Overall, at the upcoming HL-LHC and future FCC, heavy-ion UPCs provide an additional avenue for investigating the production of doubly heavy hadrons.

%%%%%%%%%%%%%%%%%%%%%%%%%%%%%%%%%%%%%%%%%%%%%%%%%%%%%%%%%%%%%%%%%%%%%
\vspace{0.5cm} {\bf Acknowledgments}
The author thanks the organisers of Hadron2025 conference for the excellent organisation of a very educating conference.
This work is supported by the National Natural Science Foundation of China (NSFC) under the Grants No. 12475083.
%%%%%%%%%%%%%%%%%%%%%%%%%%%%%%%%%%%%%%%%%%%%%%%%%%%%%%%%%%%%%%%%%%%%

\end{document}